\documentclass[12pt,preprint]{aastex}
\slugcomment{Submitted to {\it The Astrophysical Journal}}
\shortauthors{Kirkpatrick}
\shorttitle{WISE Y Dwarfs}

\usepackage{color,amsmath}

\begin{document}

\title{Discovery of the Y1 Dwarf WISE J064723.23$-$623235.5}

\author{J.\ Davy Kirkpatrick\altaffilmark{a},
Michael C.\ Cushing\altaffilmark{b},
Christopher R.\ Gelino\altaffilmark{a},
Charles A.\ Beichman\altaffilmark{a},
C.\ G.\ Tinney\altaffilmark{c,d},
Jacqueline K.\ Faherty\altaffilmark{e},
Adam Schneider\altaffilmark{b}, and
Gregory N.\ Mace\altaffilmark{a,f}
}

\altaffiltext{a}{Infrared Processing and Analysis Center, MS 100-22, California 
    Institute of Technology, Pasadena, CA 91125, USA; davy@ipac.caltech.edu}
\altaffiltext{b}{Department of Physics and Astronomy, MS 111, University of Toledo, 2801 W. Bancroft St., Toledo, OH 43606-3328, USA}
\altaffiltext{c}{School of Physics, University of New South Wales, Sydney, NSW 2052, Australia}
\altaffiltext{d}{Australian Centre for Astrobiology, University of New South Wales, Sydney, NSW 2052, Australia}
\altaffiltext{e}{Department of Astronomy, University of Chile, Camino El Observatorio 1515, Casilla 36-D, Santiago, Chile}
\altaffiltext{f}{Department of Physics and Astronomy, UCLA, 430 Portola Plaza, Box 951547, Los Angeles, CA, 90095-1547, USA}

\begin{abstract}

We present the discovery of a very cold, very low mass, nearby brown dwarf using data from the NASA {\it Wide-field Infrared Survey Explorer} (WISE). The object, WISE J064723.23$-$623235.5, has a very red WISE color of W1$-$W2 $>$ 3.77 mag and a very red {\it Spitzer Space Telescope} color of ch1$-$ch2 = 2.82$\pm$0.09 mag. In $J_{MKO}-$ch2 color (7.58$\pm$0.27 mag) it is one of the two or three reddest brown dwarfs known. Our grism spectrum from the {\it Hubble Space Telescope} ({\it HST}) confirms it to be the seventeenth Y dwarf discovered, and its spectral type of Y1$\pm$0.5 makes it one of the four latest-type Y dwarfs classified. Astrometric imaging from {\it Spitzer} and {\it HST}, combined with data from WISE, provides a preliminary parallax of $\pi = 115\pm12$ mas (d = 8.7$\pm$0.9 pc) and proper motion of $\mu = 387\pm25$ mas yr$^{-1}$ based on 2.5 years of monitoring. The spectrum implies a blue $J-H$ color, for which model atmosphere calculations suggest a relatively low surface gravity. The best fit to these models indicates an effective temperature of 350-400K and a mass of $\sim$5-30 M$_{Jup}$. Kinematic analysis hints that this object may belong to the Columba moving group, which would support an age of $\sim$30 Myr and thus an even lower mass of $<$2 M$_{Jup}$, but verification would require a radial velocity measurement not currently possible for a $J=22.7$ mag brown dwarf.

\end{abstract}

\section{Introduction}

Why is the discovery of a single new Y dwarf important? First, promptly announcing new Y dwarf discoveries means that the astronomical community can perform various follow-up investigations while the {\it Spitzer} and {\it Hubble Space Telescopes} are still active. Ground-based observatories are hindered by the extreme faintness of these objects below 2.5 $\mu$m (see figure 12 of \citealt{kirkpatrick2012}) and by the opaqueness of Earth's atmosphere at thermal and mid-infrared wavelengths where these objects emit most of their light. Space-based facilities do not suffer from these limitations and can be used, for example, to obtain accurate distances via precision astrometry (\citealt{beichman2013,marsh2013,dupuy2013}), to search for transiting exoplanets having brown dwarfs as their host ``suns'' (\citealt{triaud2013}), and to probe the evolution of features in the cloud deck via photometric monitoring (\citealt{buenzli2012,apai2013,heinze2013}). Unfortunately, these orbiting facilities are nearing the end of their lifetimes, so quick publication is critical.

Second, only sixteen Y dwarfs are known (\citealt{cushing2011,kirkpatrick2012,tinney2012,liu2012,cushing2013}), along with another assumed to be a Y dwarf but still lacking spectroscopic confirmation (\citealt{luhman2011}). When in the realm of small number statistics, the discovery of a single, new object in the Solar Neighborhood is still significant, as this has a substantial impact on our understanding of the Y dwarf space density. An accurate space density is crucial because it is these coldest objects that can reveal the low-mass cutoff for star formation (see, for example, the simulations in \citealt{burgasser2004}). Gaining an accurate count of the nearby Y dwarf population will enable us to study how abruptly, or not, the observed space density below 300K drops (Figure 14 of \citealt{kirkpatrick2012}), which in turn will shed light on whether we have found the end of star formation or whether far colder Y dwarfs, beyond the reach even of NASA's {\it Wide-field Infrared Survey Explorer} (WISE; \citealt{wright2010}), are expected. Stated another way, does the low apparent space density of the latest Y dwarfs argue that we have already found star formation's lowest-mass by-products; do these objects represent another population entirely, such as rogue planets (\citealt{beichman2013}, \citealt{sumi2011}); or does the diversity of the spectral and photometric properties of the Y dwarfs merely reflect the empirical dispersion of cold, chemically complex atmospheres?

To this end, we present the discovery of a new Y dwarf whose follow-up data indicate that it is one of the coldest and lowest mass solivagant objects yet found outside our solar system.

\section{Discovery}

\begin{figure}
\epsscale{0.8}
\figurenum{1}
\plotone{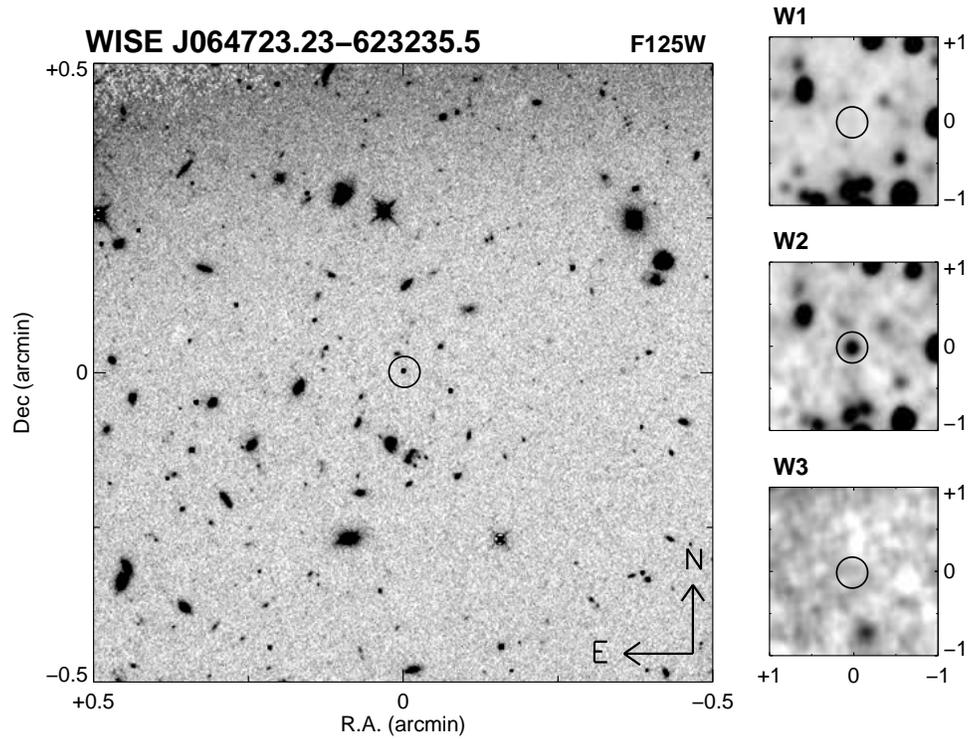}
\caption{Finder chart for WISE 0647$-$6232. The large image on the left is a 1$\times$1 arcminute zoom taken in the F125W band of $HST$/WFC3; the Y dwarf is circled. Along the right are 2$\times$2 arcminute images from WISE in bands W1, W2, and W3 with the location of the Y dwarf shown by the circle. In all images, north is up and east is to the left.
\label{finder}}
\end{figure}

Located in the constellation of Pictor, WISE J064723.23$-$623235.5 (hereafter abbreviated as WISE 0647$-$6232) was first imaged by WISE on 09 May 2010 and uncovered as a very cold brown dwarf candidate on 17 Jun 2010 after preliminary data processing. The goal of this candidate selection, described in detail in \cite{kirkpatrick2011}, was to identify brown dwarf candidates of type T6 or later by searching the database for sources having a W1-W2 color greater than 2.0 mag (see Figure 1 of \citealt{kirkpatrick2011} or Figure 1 of \citealt{mace2013a}). WISE 0647$-$6232 easily satisfies this criterion with its value of W1$-$W2 $>$ 3.77 mag (WISE All-Sky Source Catalog, \citealt{cutri2012}; see Table~\ref{photometry}). The lack of a detection at other WISE bands as well as at the shorter wavelengths imaged by the Two Micron All-Sky Survey (2MASS; \citealt{skrutskie2006}) and the Digitized Sky Survey, further cemented this object as an excellent cold brown dwarf candidate. A finder chart is shown in Figure~\ref{finder}.

\section{Follow-up Observations}

\subsection{Spitzer/IRAC}\label{Spitzer}

Because WISE 0647$-$6232 is not detected in the W1 band, it was targeted for follow-up during the Cycle 7 {\it Spitzer Space Telescope} program 70062 (PI: Kirkpatrick) using the Infrared Array Camera (IRAC; \citealt{fazio2004}). Warm IRAC employs 256$\times$256-pixel detector arrays to image a 5$\farcm$2$\times$5$\farcm$2 (1$\farcs$2 pixel$^{-1}$) field of view at wavelengths very similar to the two shortest wavelengths of WISE. The 3.6 and 4.5 $\mu$m IRAC channels (hereafter referred to as ch1 and ch2, respectively) can probe much deeper than those of WISE, thus providing a definitive ch1$-$ch2 color for cases where the W1$-$W2 value is only a lower limit. It should be noted, however, that the ch1 band is shifted redward relative to the W1 band, the latter of which is better optimized to fall in the CH$_4$ absorption trough in cold brown dwarfs (see Figure 2 of \citealt{mainzer2011}). As a result, the ch1 magnitude will necessarily be brighter than the W1 magnitude for a cold brown dwarf, and consequently the ch1$-$ch2 color will be bluer than the corresponding W1$-$W2 color. 

Observations in each channel used a five-position, medium-scale ``cycling'' dither script, with thirty-second integrations taken at each dither position. As described in \cite{kirkpatrick2011}, we measured photometry directly from the post-basic calibrated data (pBCD) mosaics produced by the {\it Spitzer} Science Center's IRAC Pipeline, software version S18.18. We employed an aperture of four-pixel radius to measure photometry of the source, with an annulus of radius 24 to 40 pixels to measure the background. The pixel scale of the mosaics is 0$\farcs$6 pixel$^{-1}$. We applied aperture corrections as listed in Table 4.7 of the IRAC Instrument Handbook$\footnote{http://ssc.spitzer.caltech.edu/irac/iracinstrumenthandbook}$. Our measurement of ch1$-$ch2 = 2.82$\pm$0.09 mag (Table~\ref{photometry}) strongly suggests that this is a very cold Y-type brown dwarf, as shown by the correlation of ch1$-$ch2 color versus spectral type in figure 11 of \cite{kirkpatrick2011}. We note that \cite{griffith2012} published $Spitzer$ photometry of this source using the same program 70062 data set but using a different reduction algorithm and found a similar color of ch1$-$ch2 = 2.87$\pm$0.10 mag.

We have continued to use {\it Spitzer}/IRAC to monitor the astrometry of this source at ch2. After our initial ch1 and ch2 observations in program 70062, we observed the source during the same program several months later but only in ch2. In Cycle 8 program 80109 (PI: Kirkpatrick) we used the same data acquisition methodology employed for program 70062 and observed the source at an additional epoch. In Cycle 9-10 program 90007 (PI: Kirkpatrick) we altered the methodology to better optimize the astrometric potential of the data by employing a nine-point random dither pattern of medium scale with thirty-second exposures per dither. The source has been reobserved multiple times using this modified mode. Table~\ref{astrometry} lists the observation dates for all of our observations in programs 70062 and 80109 and for the observations so far acquired in program 90007.

\subsection{Magellan/PANIC}

To prioritize this object for spectroscopic follow-up, we obtained $J$- and $H$-band images on 25 Nov 2010 (UT) using Persson's Auxiliary Nasmyth Infrared Camera (PANIC; \citealt{martini2004}) at the 6.5m Magellan Baade Telescope at the Las Campanas Observatory in Chile. PANIC has a 1024$\times$1024 HAWAII array with a field of view of 2\arcmin$\times$2\arcmin\ (0$\farcs$125 pixel$^{-1}$). The $J$ and $H$ bands are on the LCO filter system. For reference, transmission curves for the filters used in PANIC are shown in Figure 2 of \cite{hamuy2006} and can be compared to the transmission curves shown in Figure 4 of \cite{bessell2005}. The half-power points of the $H$ filter cut-ons and cut-offs are similar between the PANIC and MKO filters, but the $J_{LCO}$ filter is somewhat broader than $J_{MKO}$, as can also be seen in Figure 5 of \cite{stephens2004}.

Data were reduced using custom Interactive Data Language (IDL) routines further described in \cite{kirkpatrick2011}. Briefly, each image was first corrected for non-linearity. A nightly dark frame, formed from the median of many individual dark frames, was subtracted from each twilight flat frame. Each flat frame was then scaled to a common flux level and medianed to form a flat field image. After sky subtraction, target frames were flat fielded and corrected for optical distortions. These source frames were then aligned and combined using a median to produce a final mosaic. Three 2MASS stars in the field of view provided both the astrometric and photometric calibration standards. Standard aperture photometry was measured using an aperture of radius 0$\farcs$75. Despite total exposure times of 900 s at both $J$ and $H$, WISE 0647$-$6232 was not detected in either band. Table~\ref{photometry} lists the resulting 4$\sigma$ (99.99\% confidence) limits of $J_{LCO} > 23.0\pm0.1$ mag and $H_{LCO} > 21.7\pm0.3$ mag, for which the quoted errors are dominated by the uncertainty in the measured photometric zero-points. We have synthesized $J$- and $H$-band photometry for both the LCO and MKO-NIR systems using a spectrum of the T9 dwarf UGPS J072227.51$-$054031.2 and find that the $J_{LCO}$ value is 0.24 mag fainter than $J_{MKO}$, whereas $H_{LCO}$ is brighter than $H_{MKO}$ by only 0.02 mag. Assuming that WISE 0647$-$6232 has similar offsets, then the PANIC limits can be interpreted as $J_{MKO} > 22.8\pm0.1$ and $H_{MKO} > 21.7\pm0.3$ mag.

\subsection{Magellan/FourStar}

The object was observed with the FourStar infrared camera (\citealt{persson2008}) on 15 Jan 2013 (UT) and 23 Mar 2013 (UT) at the 6.5m Magellan Baade Telescope. Observations were performed in the methane-sensitive $J3$ and $J2$ filters in January, and again in the $J3$ filter in March (when the seeing was 0$\farcs$39). Exposure times were 2568s in each filter on each night. As illustrated in Figure 1 of \cite{tinney2012}, the $J2$ filter is centered on a region strongly suppressed by methane absorption in cold brown dwarfs, while the $J3$ filter encompasses the $J$-band opacity hole. The filters can be thought of as $J$-band analogs of the CH$_4$s and CH$_4$l filters used at $H$-band in \cite{tinney2005}. Observing procedures, data reduction, and photometric calibration are identical to those described in \cite{tinney2012}. The photometric measurements are given in Table~\ref{photometry}.

\subsection{Magellan/FIRE}

\begin{figure}
\epsscale{0.9}
\figurenum{2}
\plotone{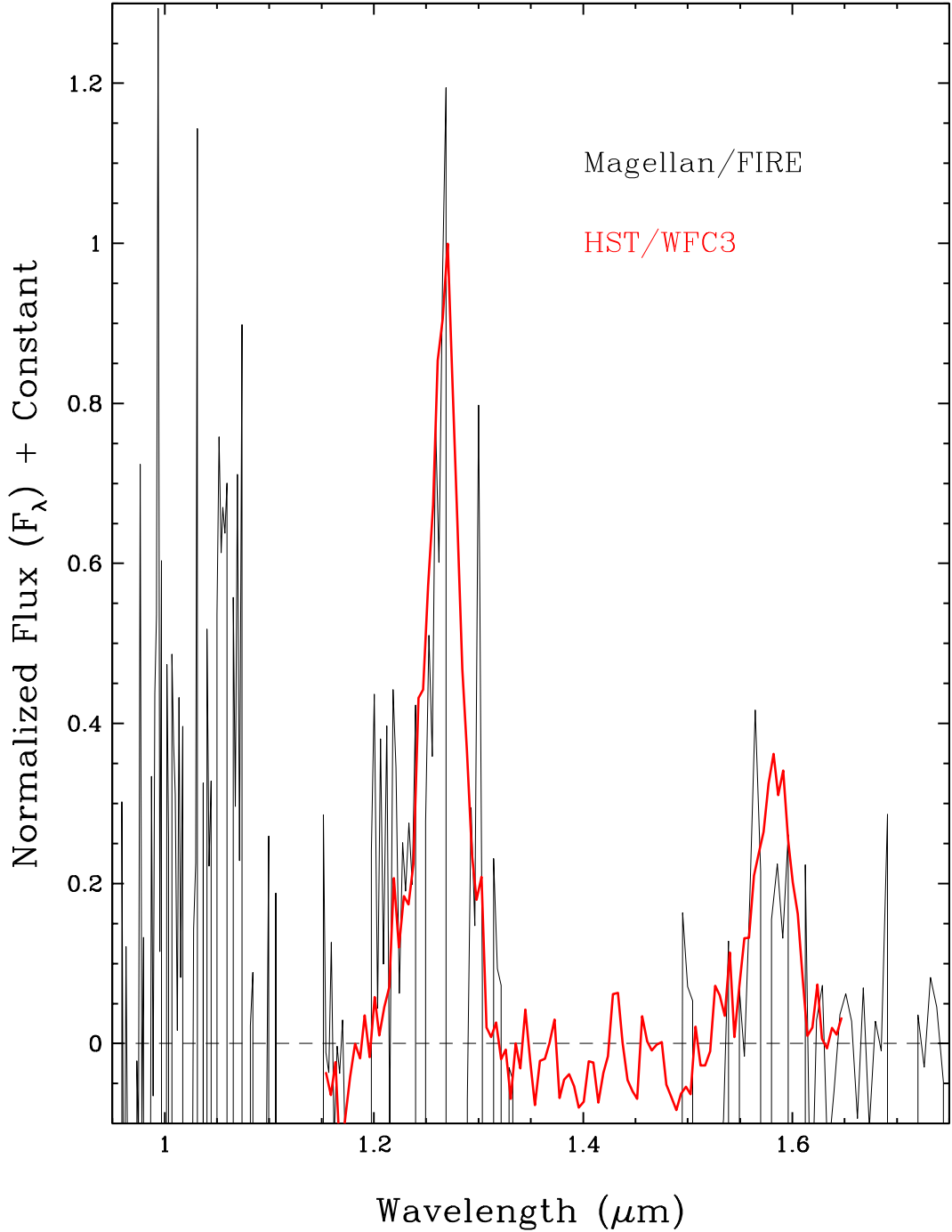}
\caption{The Magllan/FIRE spectrum of WISE 0647$-$6232 (black line) overplotted with the HST/WFC3 spectrum of WISE 0647$-$6232 (heavy red line) from Figure~\ref{spectrum}. The WFC3 spectrum has been normalized to unity at 1.28 $\mu$m and the FIRE spectrum normalized roughly to match.
\label{spectrum2}}
\end{figure}

On 24 Mar 2013 (UT), during a night of exceptional seeing ($<$0$\farcs$5) at Las Campanas, we attempted a spectroscopic observation of WISE 0647$-$6232 using the 6.5m Magellan Baade telescope with the FIRE spectrograph (\citealt{simcoe2013}) in long-slit prism mode, which gives a resolving power of $R = 400$. Four 600s integrations were taken, and these were nodded along the 0$\farcs$6 slit in an ABBA pattern. The airmass range was 1.51 to 1.67. Data processing and analysis procedures followed those outlined for the Y dwarf WISE J163940.83$-$684738.6 in \cite{tinney2012}, with the only difference being that a standard extraction was used rather than that paper's detailed removal of the contamination of a nearby star. Flux and telluric calibration was performed using an observation of the A0 V star HD 51139 obtained at a similar airmass. Figure~\ref{spectrum2} shows the resulting spectrum for WISE 0647$-$6232. Although the quality of these data is too low to permit accurate classification, they are still useful as confirmation that the source is a late-type brown dwarf because the characteristically narrow opacity holes at $Y$, $J$, and $H$ bands are the only wavelength regions where flux is detected. These data also demonstrate what is possible (at $J \approx$ 23 mag) in ideal atmospheric conditions using a large-aperture ground-based telescope with a suitable low-resolution infrared spectrograph and 40m of on-source integration. 

\subsection{HST/WFC3}\label{HST}

Higher signal-to-noise spectroscopy was acquired using the Wide Field Camera 3 (WFC3) onboard {\it HST} as part of Cycle 20 program 12970 (PI: Cushing). WFC3 employs a 1024$\times$1024 HgCdTe detector with a plate scale of 0$\farcs$13 pixel$^{-1}$ to image a field of view of 123\arcsec$\times$126\arcsec. We first obtained pre-images of each field so that we could determine specific roll angles for the slitless grism images. This was done to ensure that the object's spectrum would not be contaminated by those of field sources. Our pre-image for WISE 0647$-$6232 was acquired on 11 Feb 2013 (UT) in the F125W filter, which is a close approximation to the $J$ band. Integrations at four different dither positions were taken in a single orbit with the SPARS50 sampling sequence and thirteen MULTIACCUM samples for a total exposure time of 2412 s. Using this image, we computed roll angles for the grism observations. 

These grism observations, acquired on 13/14 May 2013 (UT), consisted of direct images in F125W (SPARS25 sampling with eight to twelve MULTIACCUM samples) at three dithered positions in addition to dispersed images in G141 (SPARS100 sampling with thirteen MULTIACCUM samples) at four dithered positions. The total integration time was 734 s for the direct images and 9624 s for the dispersed images. The G141 grism was used for the spectroscopic observations because it covers the 1.1-1.7 $\mu$m range (resolving power of $R$$\approx$130) that samples the diagnostic $J$- and $H$-band peaks in the spectrum of cold brown dwarfs.

\begin{figure}
\epsscale{0.8}
\figurenum{3}
\plotone{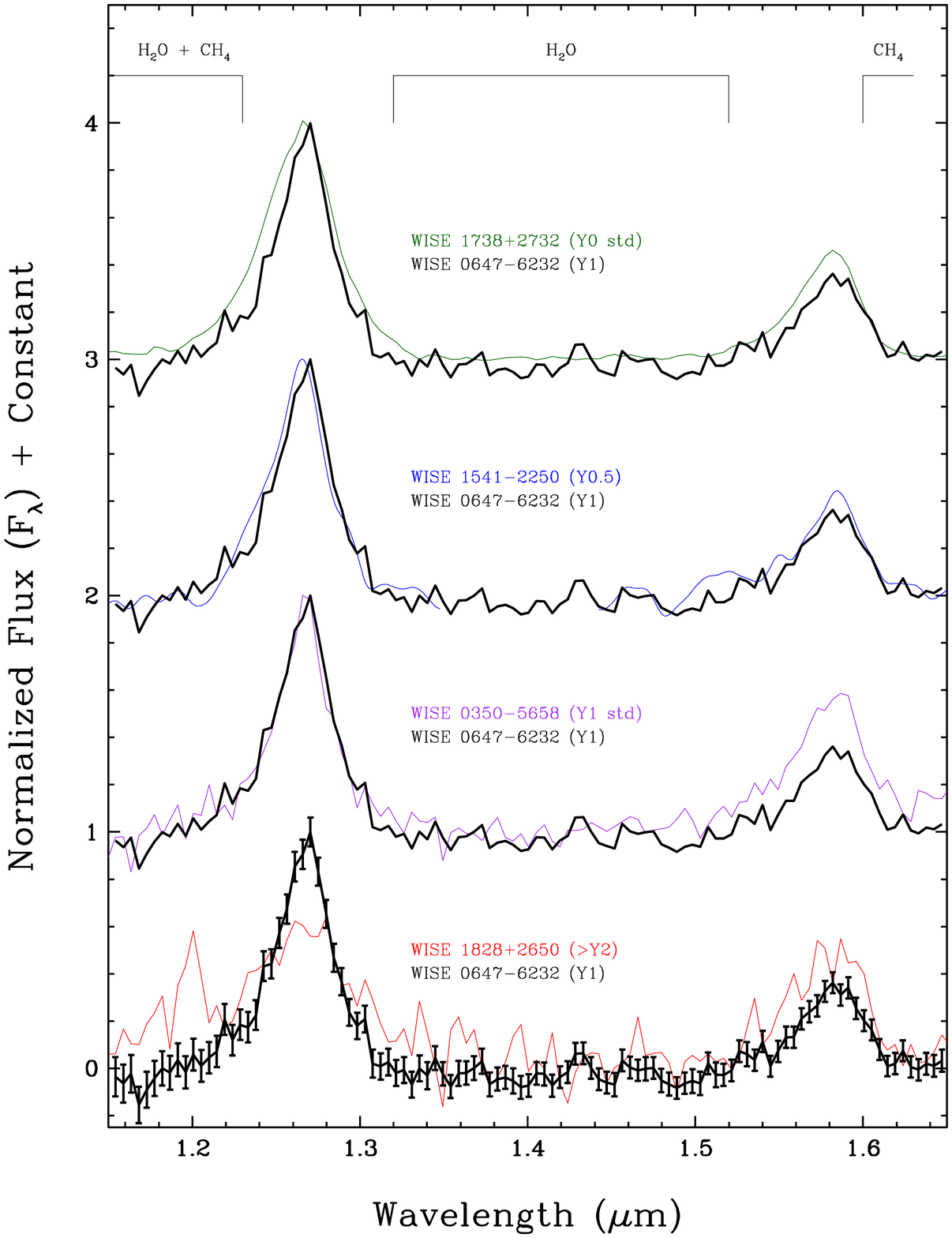}
\caption{Spectrum of WISE 0647$-$6232 (heavy black line) compared to the Y0 standard WISE 1738+2732 (\citealt{cushing2011}; light green line), the Y0.5 dwarf WISE 1541$-$2250 (\citealt{kirkpatrick2012}; light blue line), the tentative Y1 standard WISE 0350$-$5658 (\citealt{kirkpatrick2012}; light purple line), and the $>$Y2 dwarf WISE 1828+2650 (\citealt{cushing2011}; red line). All spectra, except for that of WISE 1828+2650, have been normalized to one at the $J$-band peak, and integer offsets added to the y-axis to separate the spectra vertically except when overplotting was intended. Error bars for the WISE 0647$-$6232 data are shown on the bottom spectrum.
\label{spectrum}}
\end{figure}

Photometry was measured on the drizzled images, and magnitudes were measured on the Vega system. Photometric reductions followed the recipe described in \cite{cushing2011}. Our measured magnitude from these images\footnote{These images, rather than the deeper ones taken for the pre-imaging observations from 11 Feb 2013, were used for the photometry because the earlier data have non-uniform backgrounds.} is given in Table~\ref{photometry}. Spectroscopic reduction generally followed the recipe outlined in \cite{cushing2011}. However, a second background subtraction was required, so a two-dimensional surface was fit after masking pixels containing flux from the target, and this was subtracted from the drizzled postage stamp around the target. The spectrum was then extracted using custom IDL code using a 1\arcsec\ aperture. Flux calibration was accomplished using the sensitivity curves provided by the Space Telescope Science Institute (STScI). The final flux-calibrated spectrum is shown in Figure~\ref{spectrum}. This spectrum has been used to synthesize a $J-H$ color using the MKO $J$ and $H$ filter profiles as described in \cite{cushing2005}, resulting in $(J-H)_{MKO}$ = $-$0.75$\pm$0.10 mag (Table~\ref{photometry}). The {\it HST} spectrum cuts off shortward of the long-wavelength edge of the $H_{MKO}$ bandpass, so we have assumed that the flux is zero here, as is the case for Y dwarfs and late-T dwarfs with spectra covering the full range. The uncertainty was derived by measuring the color for a suite of a Monte Carlo simulated spectra derived from the measured errors in each spectral bin.

\section{Determining Spectral Type}

The $J3-J2$ color of $-$1.21$\pm$0.20 mag from Table~\ref{photometry} falls blueward of the $J3-J2 = -0.5$ mag limit that \cite{tinney2012} cite as strong evidence for significant methane absorption. The Magellan/FIRE spectrum confirms this, and strongly hints that the object is a late-T or early-Y dwarf. The $J3-$W2 color of 7.19$\pm$0.12 mag (Table~\ref{photometry}) implies a type of $\sim$Y1 using the $J3-$W2 versus spectral type relation shown in Figure 4 of \cite{tinney2012}.

The actual classification of WISE 0647$-$6232 is determined via by-eye comparison of the {\it HST}/WFC3 spectrum to the Y0 standard WISEP J173835.52+273258.9 of \cite{cushing2011} and the tentative Y1 standard WISE J035000.32$-$565830.2 of \cite{kirkpatrick2012}, as shown in Figure~\ref{spectrum}. As further illustration, the spectrum is also compared to the Y0.5 dwarf WISE J154151.65$-$225024.9 and the $\ge$Y2 dwarf WISE J182831.08+265037.7, whose classifications also come from \cite{kirkpatrick2012}. As the figure shows, the $J$-band peak more closely matches the Y1 standard than it does the Y0 standard, and overall is also narrower than the $J$-band peak of the Y0.5 dwarf. Given the close match to the overall $J$-band spectrum of the Y1 standard, we classify WISE 0647$-$6232 as a Y1 dwarf, where this classification carries the usual uncertainty of a half subclass. We have also computed the J-narrow index of \cite{mace2013a} using the methodology described in that paper and find J-narrow = 0.69$\pm$0.07. This value is lower than all of the normal Y0 dwarfs in Table 7 of \cite{mace2013a} and is most like the J-narrow = 0.68$\pm$0.10 value of WISE J035000.32$-$565830.2, also classified as Y1. Thus, this index quantifies what can be seen qualitatively by eye.

\section{Determining Parallax and Proper Motion}

We have measured preliminary trigonometric parallax and proper motion values using our available imaging. The resultant astrometry is listed in Table~\ref{astrometry} and discussed further below.

The {\it HST} images consist of the F125W images discussed in section \ref{HST} as well as F105W images that will be discussed in a future paper (Schneider et al., in prep.). The STScI's ``AstroDrizzle'' pipeline  was used to process these data to produce final mosaicked images. The pipeline corrects for the geometric distortion of the WFC3-IR camera to a level estimated to be $<$ 8 mas (\citealt{kozhurina2009}), which is of the same order or less than  the extraction uncertainties of the faint target.  Sources were extracted using the Gaussian-fitting IDL FIND routine to determine centroid positions. The full width at half maximum (FWHM) for these undersampled data is $\sim$ 2 pixels (0$\farcs$26), which is consistent with STScI analyses (\citealt{kozhurina2009}). The images were placed onto a common ($\alpha,\delta$) origin using averaged positions for over 300 stellar-like objects within 45\arcsec\ of the Y dwarf. For nine objects seen in all epochs and brighter than F125W = 20 mag, the single axis positional dispersion is 4 mas. For nine unconfused, stellar objects in the range 23.0 $<$ F125W $<$ 23.8 mag straddling the F125W = 23.4 mag brightness of WISE 0647$-$6232, the single-epoch positional repeatability is 15 mas. These measures of the positional dispersion appear, however, to be conservative since the repeatability of the one F125W and two F105W observations taken over the two day period in May 2013 (MJD $\approx$ 56426) is only 4 mas. We therefore take a compromise value of 10 mas for the single-epoch {\it HST} uncertainties.

Seven separate {\it Spitzer} observations -- one at ch1 and six at ch2 -- were taken as described in section \ref{Spitzer}. We analyzed pBCD mosaics from the {\it Spitzer} Science Center (SSC) to make astrometric measurements. We combined all epochs of ch1 and ch2 to a common reference by averaging together the positions of 85 bright sources within $\sim$60\arcsec\ of WISE 0647$-$6232 and calculating small offsets from one epoch to the next to register all frames to the average value. The largest offsets were of order 200 mas but were typically much smaller, around 50 mas.  The common origin for these 85 sources was determined with a precision of $(\Delta\alpha=-1.2\pm5,\Delta\delta=0.2\pm0.2)$ mas. For a single sighting of a source brighter than ch2 $<$ 17 mag, the positional repeatability is 60 mas and increases as expected at fainter magnitudes. These values are less than the quoted 100 mas optical distortions cited by the SSC\footnote{http://irsa.ipac.caltech.edu/data/SPITZER/docs/irac/iracinstrumenthandbook/26/}, in part because we are confining our observations to a small central part of the IRAC arrays. 

As a final step in preparing the positional data for astrometric analysis, we shifted the positions onto a common reference frame by comparing twelve objects seen in common in the 60\arcsec\ field around the brown dwarf, taking  care to exclude sources that were obviously extended or double as seen in the higher-resolution {\it HST} images. After correction of $(\alpha,\delta)$ offsets $<$100 mas, the frame pairs share a common origin with an accuracy of $(\sigma_\alpha,\sigma_\delta)=(10,15)$ mas, which is negligible compared to the extraction uncertainties of the {\it Spitzer} data.   The  {\it Spitzer}  and WISE astrometric frames (\citealt{cutri2012}) are tied to the 2MASS frame, which has an overall absolute accuracy of 80 mas (\citealt{skrutskie2006}). The absolute positional accuracy is not, however, critical to the astrometric solution described below.

\begin{figure}
\epsscale{1.0}
\figurenum{4}
\plotone{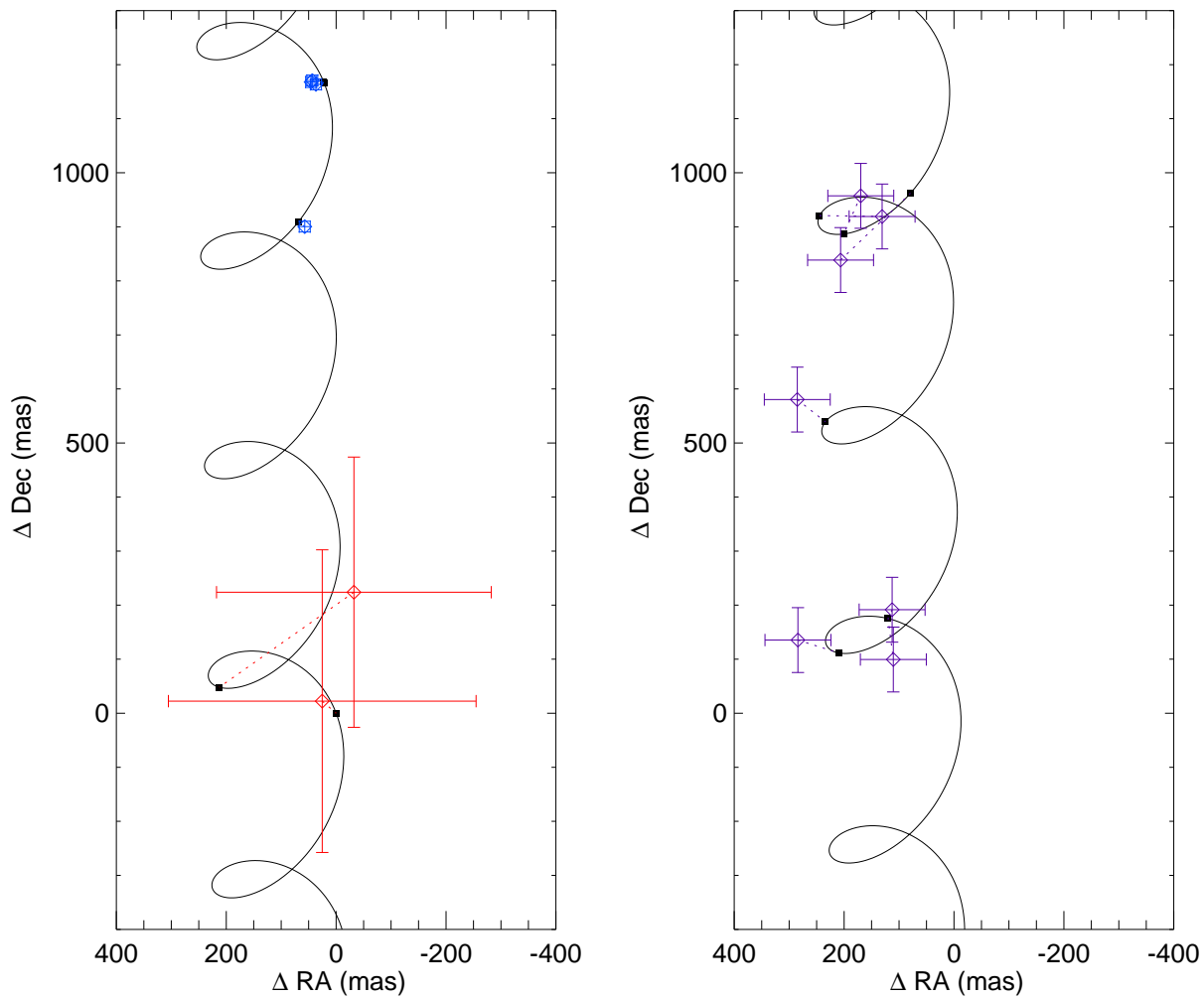}
\caption{The predicted path of WISE 0647$-$6232 on the sky (black) based on our best fit to the available astrometry. The astrometric path as seen from the Earth-orbiting observatories {\it HST} and WISE is shown on the left, and the path as seen from {\it Spitzer} in its Earth-trailing orbit is shown on the right. Black squares show the times at which the actual measurements were made, from Table~\ref{astrometry}. The measurements themselves are shown in red (WISE), blue ({\it HST}), and purple ({\it Spitzer}) with dashed lines of the appropriate color connecting them to the predictions. Motions are shown relative to the predicted position at the epoch of the earliest WISE observation.
\label{astrometric_path}}
\end{figure}

As described in \cite{beichman2013}, the astrometric observations are fitted to a model incorporating proper motion and parallax (\citealt{smart1977,green1985}):

\begin{align*}
\alpha^\prime\equiv&\, \alpha_0+\mu_\alpha(t-T_0)/cos(\delta^\prime)\\
\delta^\prime\equiv&\, \delta_0+\mu_\delta(t-T_0)\tag{1}\\
\\
\alpha(t)=&\, \alpha^\prime +\pi\Big( X(t) sin\,\alpha^\prime - Y(t) cos\,\alpha^\prime \Big)/cos\,\delta^\prime \\
\delta(t)=&\, \delta^\prime +\pi\Big( X(t) cos\,\alpha^\prime sin \,\delta^\prime  + Y(t) sin\,\alpha^\prime sin\,\delta^\prime -Z(t) cos\,\delta^\prime \Big)\tag{2} \\
\end{align*}

\noindent where $(\alpha_0, \delta_0)$ are the source position for equinox and epoch $T_0=$J2000.0,
($\mu_{\alpha},\mu_{\delta}$) are the proper motion values in the two coordinates in arcsec/yr, and $\pi$ is the annual parallax in arcsec. 
The coefficients $X(t),\ Y(t)$, and $Z(t)$ are the Cartesian coordinates, relative to the Sun, of WISE or {\it HST} in 
Earth orbit or {\it Spitzer} in its Earth-trailing orbit. 
Values of $X,Y,Z$ for the Earth-orbiting observatories are taken from the IDL ASTRO routine 
XYZ. {\it Spitzer} values of $X,Y,Z$ are obtained from the image headers provided by the SSC. Equations (1) and (2) are solved simultaneously using the {\it Mathematica} routine {\it NonLinearModelFit} incorporating appropriate uncertainties for each data point. The fit to the combined {\it HST} and {\it Spitzer} data sets yields a good solution for proper motion and parallax (Table~\ref{parallax_absmags}) with $\chi^2=16.5$ with 21 degrees of freedom; this can be compared to a model incorporating only proper motion, which yields $\chi^2=247$ with 22 degrees of freedom. A model incorporating a non-zero parallax is thus strongly preferred and the slightly low reduced $\chi^2$ for this case indicates that our estimates of the positional uncertainties are, as suggested above, conservative. The resultant parallax is 115$\pm$12 mas, which corresponds to a distance of 8.71$\pm$0.88 pc, and the resultant proper motion is 387$\pm$25 mas yr$^{-1}$. Table~\ref{parallax_absmags} lists these values along with absolute magnitudes derived from the photometry in Table~\ref{photometry}. Figure~\ref{astrometric_path} shows the motion of WISE 0647$-$6232 on the sky.

\section{Discussion}

Using the $J_{MKO} - J3 = 0.14{\pm}0.25$ mag correction derived by \cite{tinney2012} for Y dwarfs, we find that WISE 0647$-$6232 has $J_{MKO}-W2$ = 7.33$\pm$0.28 mag and $J_{MKO}-ch2$ = 7.58$\pm$0.27. The only other confirmed Y dwarf with redder colors is the enigmatic $\ge$Y2 dwarf WISE J182831.08+265037.7 (\citealt{cushing2011,kirkpatrick2012,beichman2013}), with $J_{MKO}-W2$ = 9.18$\pm$0.36 mag and $J_{MKO}-ch2$ = 9.25$\pm$0.35 mag (Table 2 of \citealt{kirkpatrick2012}). One other object\footnote{This object is more correctly referred to as L 97-3B (or GJ 3483B) because the ``WD'' prefix was originally intended only for white dwarfs. However, we use the designation WD 0806$-$661B anyway because this name is the one most commonly used.}, WD 0806$-$661B (\citealt{luhman2011}), is almost certainly a Y dwarf but lacks a confirming spectrum due to its faint apparent magnitude; it has $J_{MKO}-ch2 > 7.06$ mag (\citealt{luhman2012}), so it may also be redder than WISE 0647$-$6232. The $ch1-ch2$ color of 2.82$\pm$0.09 mag is redder than both WISE J182831.08+265037.7 (2.60$\pm$0.03 mag; \citealt{kirkpatrick2012}) and WD 0806$-$661B (2.77$\pm$0.16 mag; \citealt{luhman2011}), although not as red as that of other Y dwarfs, such as the Y1 dwarf WISE J035000.32$-$565830 (3.25$\pm$0.10 mag; \citealt{kirkpatrick2012}) and the Y0: dwarf WISE J220905.73+271143.9 (3.08$\pm$0.09 mag; \citealt{cushing2013}). As Figure 8 of \cite{kirkpatrick2012} shows, there is a significant scatter of $ch1-ch2$ colors within the Y0 and Y1 subtypes, perhaps indicating a range of varying physical conditions (gravity, metallicity, cloud properties, etc.) within each group.

\begin{figure}
\epsscale{0.9}
\figurenum{5}
\plotone{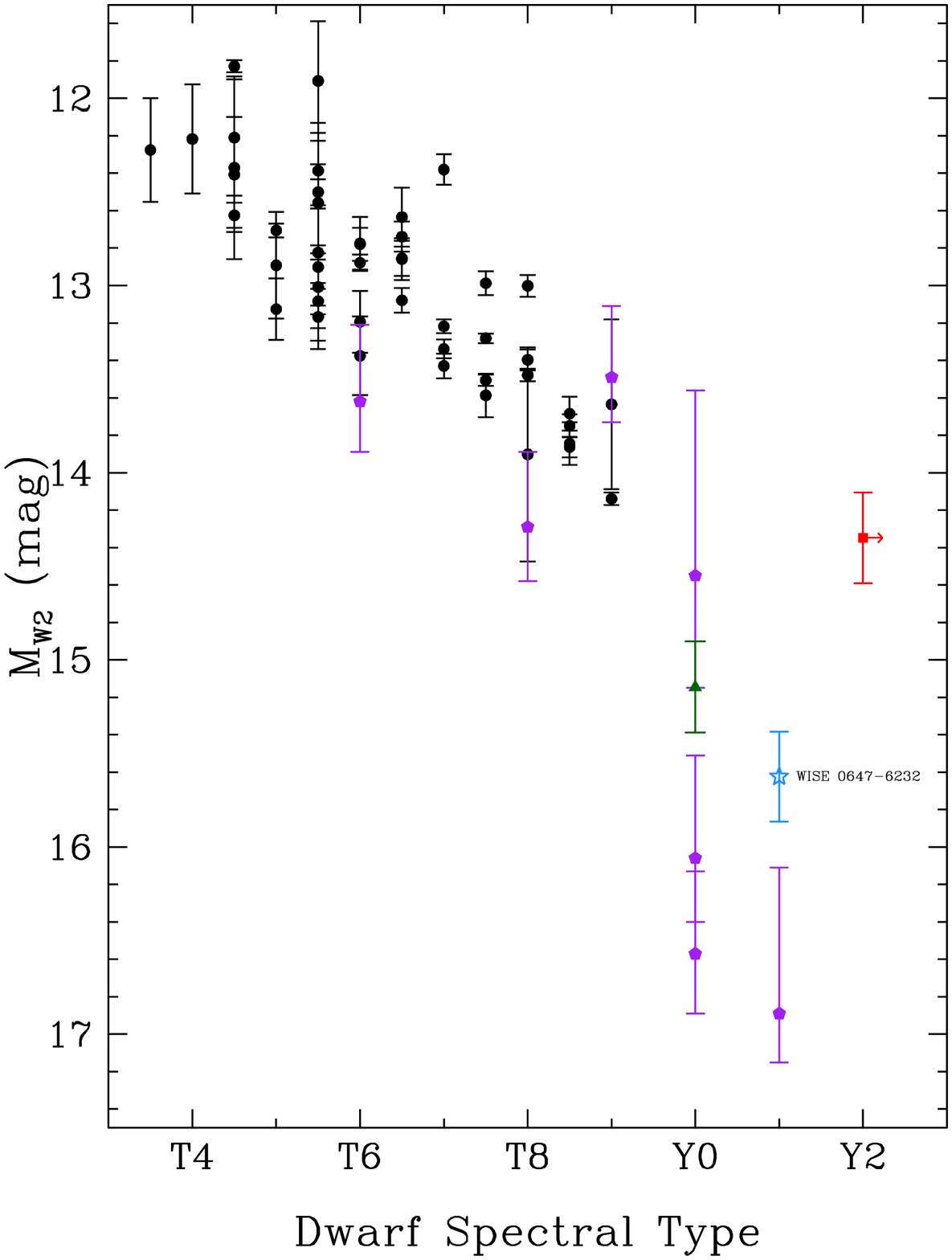}
\caption{The absolute W2 magnitude as a function of spectral type from mid-T through early-Y. Objects from the compilation of \cite{dupuy2012} are shown by black circles. Additional data from \cite{tinney2012} are shown by the green triangle, \cite{beichman2013} by the red square, and \cite{marsh2013} by the purple pentagons. The location of WISE 0647$-$6232 is shown by the open blue star. 
\label{MW2_plot}}
\end{figure}

Figure~\ref{MW2_plot} shows the trend of absolute W2 magnitude as a function of spectral type using previously published values from \cite{dupuy2012}, \cite{tinney2012}, \cite{beichman2013}, and \cite{marsh2013}. The $M_{W2}$ value for WISE 0647$-$6232 falls $\sim$1.5 mag fainter than that of the latest, well measured T dwarfs on the plot and is also fainter by $\sim$0.4 mag than the Y0: dwarf WISE J163940.83$-$684738.6 (\citealt{tinney2012}) and $\sim$1.6 mag fainter than the $\ge$Y2 dwarf WISE J182831.08+265037.7 (\citealt{beichman2013}). There are three Y0 and Y1 dwarfs from \cite{marsh2013} with fainter absolute W2 magnitudes, but all of these have error bars of $\sim$0.5 mag or higher. Thus, WISE 0647$-$6232 is one of the instrinsically dimmest, solivagant, star-like bodies yet identified outside our solar system.

\begin{figure}
\epsscale{0.8}
\figurenum{6}
\plotone{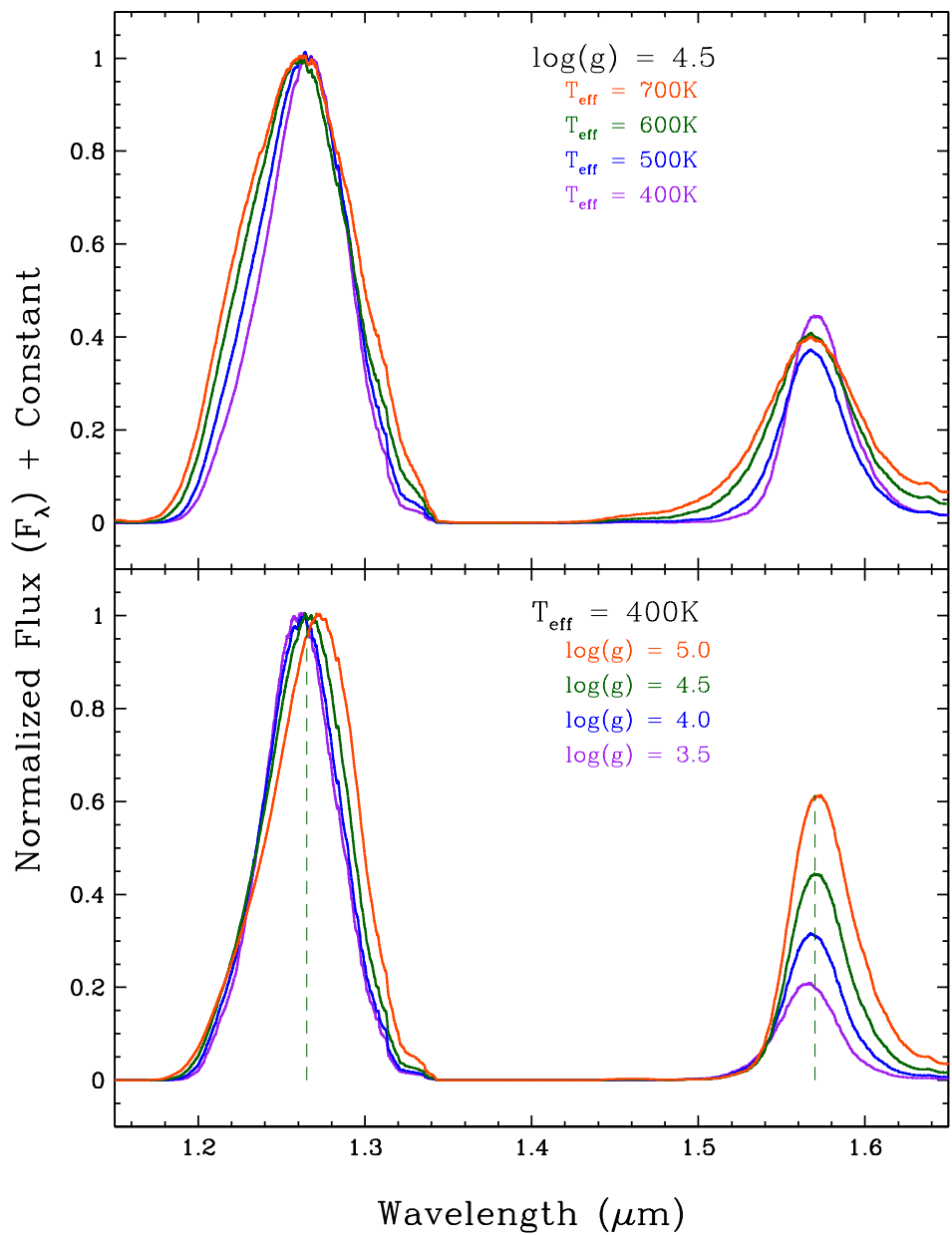}
\caption{BT-Settl models from \cite{allard2003}. All models have been smoothed to a resolution commensurate with our {\it HST}/WFC3 spectrum and normalized to one at their $J$-band peaks. (Upper panel) Models at a fixed value of log(g) = 4.5 (where g is in units of cm s$^{-2}$) for effective temperatures of 700K (orange), 600K (green), 500K (blue), and 400K (purple). (Lower panel) Models at a fixed value of T$_{eff}$ = 400K for log(g) values of 5.0 (orange), 4.5 (green), 4.0 (blue), and 3.5 (purple). Dashed green lines are used in the lower panel to denote the location in wavelength of the $J$- and $H$-band peaks for the log(g) = 4.5 model.
\label{models_fixed_logg_or_T}}
\end{figure}

What do spectroscopic models suggest about the physical parameters of WISE 0647$-$6232? Before answering this question quantitatively, it is illuminating to learn first what the models suggest about the qualitative trends. For this purpose, we use the BT-Settl models from the Lyon group (\citealt{allard2003}), which are valid down to effective temperatures of 400K (corresponding to early-Y dwarfs; \citealt{cushing2011}) and include gravitational settling of grains from the photosphere. Figure~\ref{models_fixed_logg_or_T} shows models in the range 400K $< T_{eff} <$ 700K for a fixed value of log(g) = 4.5, where g is in units of cm s$^{-2}$. These models suggest that the narrowness of the $J$- and $H$-band peaks can be used as a proxy for the brown dwarf's effective temperature. However, unlike stars on the main sequence, brown dwarfs at fixed effective temperature may span a vast array of masses, so other parameters must also be considered\footnote{With the BT-Settl model grid, as well as other model grids currently being produced, we cannot yet explore trends with [M/H] because models with halo-like metallicity have not yet been produced (see also the commentary by \citealt{mace2013b}).}. Figure~\ref{models_fixed_logg_or_T} shows the variation in spectral features over the range 3.5 $<$ log(g) $<$ 5.0 for a fixed value of $T_{eff}$ = 400K. Here, log(g) can be used as a proxy for mass. Note that even for a fixed value of $T_{eff}$ there is variation in the widths of the $J$- and $H$-band peaks. Moreover, there are also noticeable shifts in the wavelength of these peaks\footnote{Such shifts have been seen in the spectra of other Y dwarfs, most notably WISE J140518.39+553421.3, whose $H$-band peak is shifted redward (\citealt{cushing2011}) relative to other Y dwarfs, possibly indicative of a higher than average log(g) value.}, with the wavelengths of the peaks shifted slightly redward for higher values of log(g). The most noticeable effect, though, is the modulation of the $J - H$ color\footnote{An anomalously red color, such as that for WISE J182831.08+265037.7, may therefore be suggestive of a relatively high log(g) value, indicating a relatively high-mass brown dwarf (Figure 2 of \citealt{burrows2003}). However, see \cite{beichman2013} for more on the mismatch of this object's spectrum/photometry to available models over a much broader wavelength range. \cite{leggett2013} indicates that some of the mismatch between observations and models can be accounted for if the object is a binary brown dwarf.}: lower values of log(g) correspond to significantly bluer $J- H$ colors because the $H$-band peak is much more strongly muted relative to the $J$-band peak.

\begin{figure}
\epsscale{0.9}
\figurenum{7}
\plotone{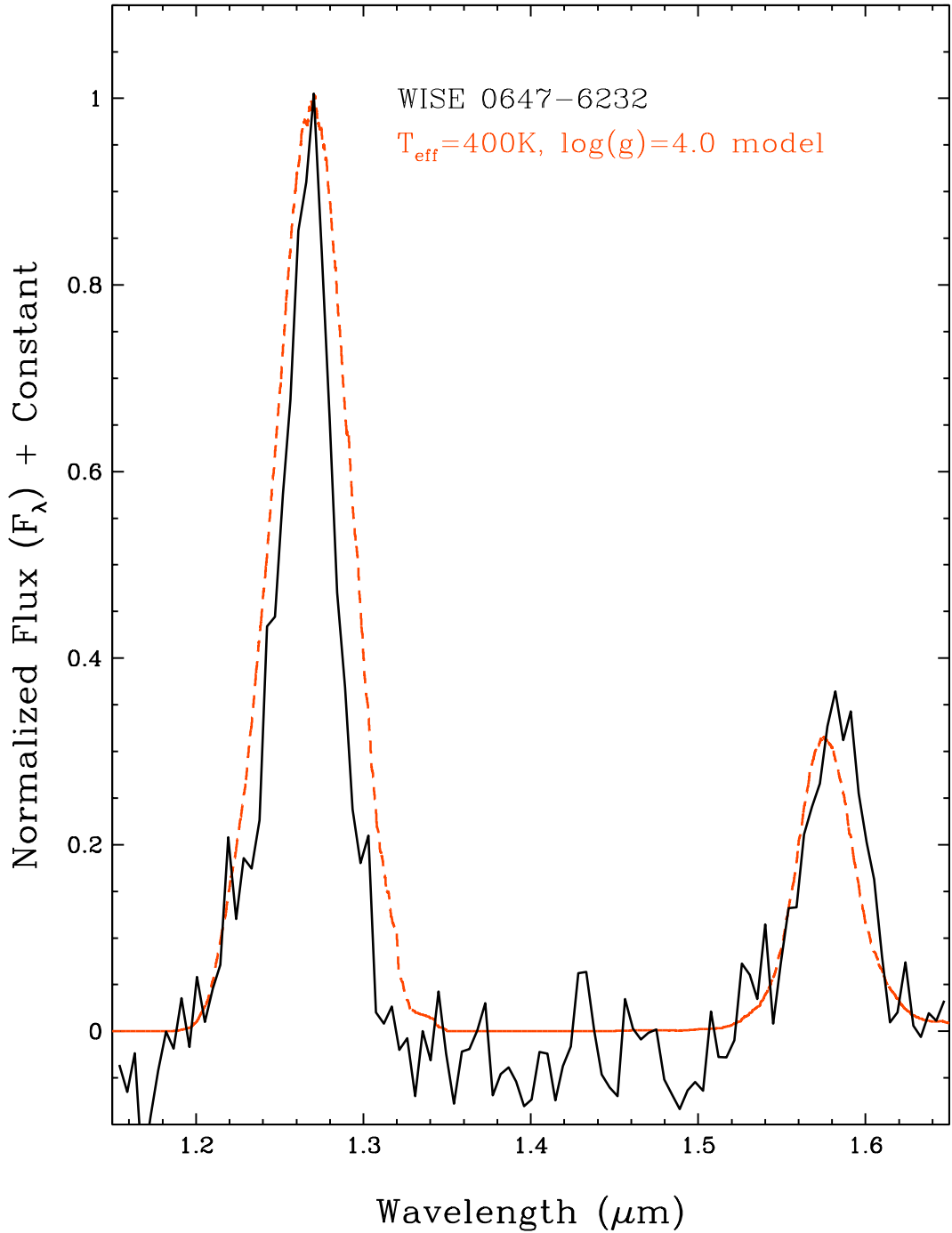}
\caption{The best fitting BT-Settl model (orange) from \cite{allard2003} overplotted on the {\it HST}/WFC3 spectrum of WISE 0647$-$6232 (black). A wavelength shift of +0.007 $\mu$m was applied to the model to bring the $J$-band peaks into agreement. This fit implies values of $T_{eff} <$ 400K and log(g) $\approx$ 4.0 for WISE 0647$-$6232.
\label{models_bestfit}}
\end{figure}

WISE 0647$-$6232 is the bluest Y dwarf in implied $J- H$ color of all the Y dwarfs shown in Figure~\ref{spectrum} and has as narrow a $J$-band peak as any Y dwarf yet observed. Qualitatively, then, we would expect WISE 0647$-$6232 to have a low effective temperature and low value of log(g) relative to other known Y dwarfs. Figure~\ref{models_bestfit}, showing the best by-eye fit to the BT-Settl models, confirms this. The best fit implies $T_{eff} <$ 400K (the $J$-band peak is narrower than any of the models in the grid) with a relatively low value of log(g) = 4.0. Evolutionary models by \cite{burrows2003} suggest a mass of $\sim$5 M$_{Jup}$ regardless of the exact value of T$_{eff}$. 

The BT-Settl models represent an earlier generation of theoretical spectra lacking sulfide and chloride clouds now known to be important at low effective temperatures (\citealt{morley2012}). What do these new models suggest about the physical parameters of WISE 0647$-$6232? Like the BT-Settl grid, the grid presented in \cite{morley2012} does not probe below an effective temperature of 400K, but these models were extended down to 300K in \cite{leggett2013}. Using Figures 4 and 5 of \cite{leggett2013}, our measured absolute values of $M_J = 22.95\pm0.35$ and $M_{W2} = 15.62\pm0.24$ mag (Table~\ref{parallax_absmags}) suggest a value of $T_{eff} \approx 350$K for WISE 0647$-$6232. The measurement of log(g) ranges from 4.0 dex (comparing to the upper panel of Figure 5 in \citealt{leggett2013}) to 5.0 dex (comparing to the upper right panel of Figure 4 in \citealt{leggett2013}). Comparison to the evolutionary models in \cite{saumon2008} suggests a mass of $\sim$5 M$_{Jup}$ -- similar to the conclusion reached using the BT-Settl spectra and \cite{burrows2003} evolutionary models -- and age of $\sim$100 Myr for the log(g) = 4 case. However, higher values of the mass and older ages are also indicated; the log(g) = 5 case suggests a mass of 30 M$_{Jup}$ and age $>$10 Gyr.

Given that WISE 0647$-$6232 is in the far southern sky, could it be a member of one of the young kinematic groups primarily located in that region of the sky? Using our measured distance and transverse velocity (Table~\ref{parallax_absmags}), we have analyzed whether there exists some value of the yet-unmeasured radial velocity that would yield U,V,W space motions suggesting membership in any of these young associations. We have checked published values of the U,V,W motions for the Argus/IC2391, TW Hydrae, Tucana/Horologium, $\beta$ Pic, AB Dor, $\eta$ Cha, Cha-Near, and Columba groups, and we find an excellent match to the Columba group if the unknown radial velocity of WISE 0647$-$6232 is later revealed to be $\sim$22 km/s. In fact, using the Bayesian technique advocated by (\citealt{malo2013}) we find a membership probability of 92.9\% without assuming a value for the radial velocty, and this rises to 99.6\% if the radial velocity is indeed 22 km/s. The age of 30 Myr for Columba would imply a mass of $<$2 M$_{Jup}$ (Figure 2 of \citealt{burrows2003}) for WISE 0647$-$6232 assuming an effective temperature range of 300-400K. This, however, cannot be reconciled with the result from spectral model fitting (above) unless the true value of log(g) for WISE 0647$-$6232 is closer to 3.5 or 3.0. 

We note, however, that the Y dwarf, at a distance of 8.7 pc, would be the closest member of Columba to the Sun; \cite{torres2008} list the high probability members, and their distances range between 35 and 189 pc. If confirmed, WISE 0647$-$6232 would be the closest known member of the group; at a distance of 8.7 pc, it falls 2.4$\sigma$ from the mean cluster distance of 82$\pm$30 pc given by \cite{torres2008}. If the list of Columba members from \cite{malo2013} is used instead, we find that the distance range is 35 to 81 pc. Similarly, the XYZ values for WISE 0647$-$6232 are 2$\sigma$ from the mean values noted in that paper. This indicates either a chance alignment with the space velocities of Columba or, if WISE 0647$-$6232 is a bonafide member, its low mass may have subjected it to kinematic heating and displaced it from the core of the moving group.

\section{Acknowledgments}

We thank our referee, Sandy Leggett, for a thoughtful and prompt report. We also thank Sergio Fajardo-Acosta for a careful reading of the manuscript.
This publication makes use of data products from WISE, which is a 
joint project of the University of California, Los Angeles, and the Jet Propulsion Laboratory (JPL)/California 
Institute of Technology (Caltech), funded by the National Aeronautics and Space Administration (NASA). 
 This work is based in part on observations made with the {\it Spitzer Space Telescope}, which is
operated by JPL/Caltech under a contract with
NASA. Support for this work was provided by NASA through an award issued to programs 70062, 80109, and 90007 by JPL/Caltech. 
This work is also based in part on observations made with the NASA/ESA {\it Hubble Space Telescope}, obtained
at STScI, which is operated by the Association of Universities for
Research in Astronomy, Inc., under NASA contract NAS 5-26555. These observations are associated with 
programs 12970. Support for these programs was provided by NASA through a grant from STScI.
This research has made use of the NASA/IPAC Infrared Science Archive,
which is operated by JPL/Caltech, under contract
with NASA. This paper includes data gathered with the 6.5m Magellan Telescopes located at Las Campanas Observatory, Chile. 
C.G.T.\ acknowledges the support of ARC grants DP0774000 and DP130102695. Australian access to the Magellan Telescopes was supported through the National Collaborative Research Infrastructure Strategy of the Australian Federal Government. Travel support for Magellan observing was provided by the Australian Astronomical Observatory. 
This paper also includes Magellan data granted by the National Optical Astronomy Observatories (Proposal ID 2010B-0184, P.I. Mainzer) through the Telescope System Instrumentation Program (TSIP). TSIP is funded by National Science Foundation.
We are also indebted to the SIMBAD database,
operated at CDS, Strasbourg, France.

\clearpage

\begin{deluxetable}{lcc}
\tablewidth{4.0in}
\tablenum{1}
\tablecaption{Photometry for WISE 0647$-$6232\label{photometry}}
\tablehead{
\colhead{Band/Color} &                          
\colhead{Value (mag)} &       
\colhead{Obs.\ date (UT)} 
}
\startdata
$J_{LCO}$\tablenotemark{a}   & $>$23.0$\pm$0.1    &  25 Nov 2010 \\
$J_{MKO}$  & 22.65$\pm$0.27                  &  ---\tablenotemark{b}\\
$J2$       & 23.69$\pm$0.17                  &  15 Jan 2013 \\
$J3$       & 22.51$\pm$0.09                  &  multi.\ date\tablenotemark{c}\\
$F125W$    & 23.44$\pm$0.09                  &  13 May 2013 \\
$H_{LCO}$\tablenotemark{d}   & $>$21.7$\pm$0.3                 &  25 Nov 2010 \\
$H_{MKO}$  & 23.40$\pm$0.29                  &  ---\tablenotemark{e}\\
ch1        & 17.89$\pm$0.09                  &  19 Sep 2010 \\
ch2        & 15.07$\pm$0.02                  &  19 Sep 2010 \\
W1         & $>$19.09\tablenotemark{f}       &  12 May 2010\tablenotemark{g} \\
W2         & 15.32$\pm$0.08                  &  12 May 2010\tablenotemark{g} \\
W3         & $>$13.49\tablenotemark{f}       &  12 May 2010\tablenotemark{g} \\
W4         & $>$9.66\tablenotemark{f}        &  12 May 2010\tablenotemark{g} \\ 
\hline
$J3-J2$    & $-$1.21$\pm$0.20                &  15 Jan 2013 \\
$(J-H)_{MKO}$&  $-$0.75$\pm$0.10             &  ---\tablenotemark{h}\\
W1$-$W2    & $>$3.77\tablenotemark{d}        &  12 May 2010\tablenotemark{g} \\
ch1$-$ch2  & 2.82$\pm$0.09                   &  19 Sep 2010 \\
\hline
$J_{MKO}-$W2 & 7.33$\pm$0.28                  &  --- \\
$J3-$W2    & 7.19$\pm$0.12                   &  --- \\
$F125W-$W2 & 8.12$\pm$0.12                   &  --- \\
$H_{MKO}-$W2 & 8.08$\pm$0.30     &  --- \\
\hline
$J_{MKO}-$ch2 & 7.58$\pm$0.27                 &  --- \\
$J3-$ch2   & 7.44$\pm$0.09                   &  --- \\
$F125W-$ch2& 8.37$\pm$0.09                   &  --- \\
$H_{MKO}-$ch2 & 8.33$\pm$0.29    &  --- \\
\enddata
\tablenotetext{a}{As explained in the text, this can be interpreted as $J_{MKO} > 22.8\pm0.1$ mag.}
\tablenotetext{b}{This magnitude is estimated from the measured $J3$ mag using the $J_{MKO} - J3 = 0.14{\pm}0.25$ mag offset measured by \cite{tinney2012} for other Y dwarfs measured in both filters.}
\tablenotetext{c}{The individual measurements are $J3$ = 22.48$\pm$0.12 mag (15 Jan 2013) and $J3$ = 22.55$\pm$0.12 mag (23 Mar 2013).}
\tablenotetext{d}{As explained in the text, this can also be interpreted as $H_{MKO} > 21.7\pm0.3$ mag.}
\tablenotetext{e}{This magnitude is estimated from the $J_{MKO}$ value derived in footnote {\it b} along with the $(J-H)_{MKO} = -0.75\pm0.10$ mag color measured from the {\it HST} spectrum.}
\tablenotetext{f}{This is the 95\% confidence upper limit from the WISE All-Sky Source Catalog.}
\tablenotetext{g}{These values from the WISE All-Sky Source Catalog are from coadded data whose individual frames span the range from 09 May 2010 to 15 May 2010 (UT).}
\tablenotetext{h}{Color measured from the {\it HST} spectrum.}
\end{deluxetable}

\clearpage

\begin{deluxetable}{lcccccc}
\tablenum{2}
\tablecaption{Astrometry for WISE 0647$-$6232\label{astrometry}}
\tablehead{
\colhead{Telescope} &                          
\colhead{Program ID\#} &                          
\colhead{Band} &       
\colhead{MJD} &
\colhead{RA} &
\colhead{Dec} &
\colhead{error} \\ 
\colhead{} &                          
\colhead{} &                          
\colhead{} &       
\colhead{} &
\colhead{(deg)} &
\colhead{(deg)} &
\colhead{(arcsec)} \\ 
}
\startdata
WISE\tablenotemark{a}      & ---   & --    &  55328.09 & 101.8467715 & -62.5432318  & 0.28 \\
$Spitzer$ & 70062 & ch1   &  55458.43 & 101.8468201 & -62.5431861  & 0.06 \\
$Spitzer$ & 70062 & ch2   &  55458.43 & 101.8468326 & -62.5432008  & 0.06 \\
WISE\tablenotemark{b}      & ---   & --    &  55519.76 & 101.8467368 & -62.5431758  & 0.25 \\
$Spitzer$ & 70062 & ch2   &  55583.05 & 101.8469351 & -62.5431916  & 0.06 \\
$Spitzer$ & 80109 & ch2   &  55892.34 & 101.8469185 & -62.5430848  & 0.06 \\
$Spitzer$ & 90007 & ch2   &  56264.46 & 101.8468637 & -62.5430184  & 0.06 \\
$Spitzer$ & 90007 & ch2   &  56329.15 & 101.8469220 & -62.5430029  & 0.06 \\
$HST$     & 12970 & $F125W$& 56334.40 & 101.8468365 & -62.5430078  & 0.03 \\
$Spitzer$ & 90007 & ch2   &  56390.09 & 101.8468609 & -62.5429880  & 0.06 \\
$HST$     & 12970 & $F105W$& 56425.18 & 101.8467858 & -62.5429180  & 0.03 \\
$HST$     & 12970 & $F125W$& 56425.91 & 101.8467827 & -62.5429213  & 0.03 \\
$HST$     & 12970 & $F105W$& 56426.97 & 101.8468040 & -62.5429180  & 0.03 \\
\enddata
\tablenotetext{a}{Astrometry taken from the coadd image created for the 4-band WISE All-Sky Release.}
\tablenotetext{b}{Astrometry taken from a coadd image created from the individual frames from the 2-band WISE Post-Cryo Release.}
\end{deluxetable}

\clearpage

\begin{deluxetable}{lr}
\centering
\tablewidth{4.0in}
\tablenum{3}
\tablecaption{Distance, Kinematics, and Absolute Magnitudes for WISE 0647$-$6232\label{parallax_absmags}}
\tablehead{
\colhead{Parameter}&
\colhead{Value}\\
}
\startdata
$\alpha$ (J2000.0 at epoch 2000.0)               &06$^h$47$^m$23$\fs$2270$\pm$0$\fs$0103\\
$\delta$ (J2000.0 at epoch 2000.0)               &$-$62$^o$32$^\prime$39$\farcs$744$\pm$0$\farcs$295\\
$\mu_\alpha$ (\arcsec\ yr$^{-1}$) &0.007$\pm$0.012\\
$\mu_\delta$ (\arcsec\ yr$^{-1}$) &0.387$\pm$0.022\\
$\pi$ (\arcsec)                  &0.115$\pm$0.012\\
Distance (pc)                    &8.7$\pm$0.9\\
V$_{tan}$ (km s$^{-1}$)           &16$\pm$2\\
reduced $\chi^2$ (w/ parallax)   &0.79\\
reduced $\chi^2$ (w/o parallax)  &11.2\\
\hline
$M_{J, MKO}$ (mag)                &22.95$\pm$0.35\\
$M_{J2}$ (mag)                   &23.99$\pm$0.28 \\
$M_{J3}$ (mag)                   &22.81$\pm$0.24 \\
$M_{F125W}$ (mag)                 &23.74$\pm$0.24 \\
$M_{H, MKO}$ (mag)               &23.70$\pm$0.36 \\
$M_{ch1}$ (mag)                  &18.19$\pm$0.24 \\
$M_{ch2}$ (mag)                  &15.37$\pm$0.23 \\
$M_{W2}$ (mag)                   &15.62$\pm$0.24 \\
\enddata
\end{deluxetable}

\clearpage

\clearpage

\clearpage

\clearpage

\clearpage

\end{document}